%% file: instab_revtex.tex
\documentstyle[fullpage,latexsym,epsf,aps]{revtex}
\newcommand{\UKcol}{{\sf UKQCD collaboration}}
\newcommand{\Expt}[1]{\langle #1 \rangle}
\newcommand{\dtau}{\delta \tau}
\newcommand{\U}{{\cal U}}
\newcommand{\Tr}{{\rm Tr}} 
\newcommand{\dDU}{\Delta \delta U}
\newcommand{\dDP}{\Delta \delta \pi}
\newcommand{\dDH}{\Delta \delta H}
\newcommand{\fm}{\mathop{\rm fm}}
\begin{document}
\title{ 
\hfill\begin{minipage}{0pt}\scriptsize \begin{tabbing}
\hspace*{\fill} Edinburgh 2000/12\\ 
\hspace*{\fill} LTH  455\\
\hspace*{\fill} hep-lat/xxx
\end{tabbing} 
\end{minipage}\\[8pt]  
Instability in the Molecular Dynamics Step of Hybrid \\
Monte Carlo in Dynamical Fermion Lattice QCD Simulations
}

\author{\frenchspacing B\'alint~Jo\'o\footnote{Current address: Department of Physics and Astronomy, University of Kentucky, Lexington KY 40506-0055, U.S.A}, Brian~Pendleton}
\address{Department of Physics and Astronomy, 
	The University of Edinburgh, \\
	The King's Buildings, Edinburgh EH9 3JZ,
	Scotland, UK}

\author{\frenchspacing Anthony~D.~Kennedy}
\address{The Maxwell Institute and Department of Physics and Astronomy, 
	The University of Edinburgh, \\
	The King's Buildings, Edinburgh EH9 3JZ,
	Scotland, UK}

\author{\frenchspacing Alan~C.~Irving}
\address{Theoretical Physics Division,
	 Department of Mathematical Sciences, \\ 
	 University of Liverpool, 
	PO Box 147, Liverpool L69 3BX, UK}

\author{\frenchspacing James~C.~Sexton}
\address{School of Mathematics, Trinity College, \\
	 Hitachi Dublin  Laboratory and
	 Center for Supercomputing in Ireland (CSI),\\
	 Dublin 2, Ireland}
 
\author{\frenchspacing Stephen~M.~Pickles}
\address{Computer Services for Academic Research (CSAR) \\
	The University of Manchester, Oxford Road, Manchester M13 9FL, UK}

\author{\frenchspacing Stephen~P.~Booth} 
\address{Edinburgh Parallel Computing Centre (EPCC) \\
The University of Edinburgh, Edinburgh EH9 3JZ, Scotland, UK}

\author{\UKcol{}}
\maketitle 
\input{body.tex}

\end{document}

%% file: body.tex
\begin{abstract}
We investigate instability and reversibility within 
Hybrid Monte Carlo simulations using a non-perturbatively improved Wilson
action. We demonstrate the onset of instability as tolerance parameters
and molecular dynamics step sizes are varied. 
We compare these findings with theoretical
expectations and present limits on simulation
parameters within which a stable and reversible algorithm is
obtained for physically relevant simulations. 
Results of optimisation experiments 
with respect to tolerance parameters are also
presented.\\
Pacs Numbers: 12.38.Gc, 11.15.Ha, 02.70.Lq
\end{abstract}
\newpage
\section{Introduction}
Hybrid Monte Carlo (HMC) \cite{AdkBjp}
remains the most widely used algorithm for
lattice QCD computations with dynamical fermions.  
In such computations, trial configurations are produced
by integrating the Hamiltonian equations of motion from an initial
configuration for some fictitious molecular dynamics (MD) time
$\tau$. Configurations are then accepted or rejected by subjecting the
energy change $\delta H$ along a trajectory to a Metropolis
\cite{Metropolis} accept/reject step.

It has been observed \cite{JansenLiu,AdkHorvath} that the equations of
motion in the MD evolution of such an algorithm are chaotic in the
case of QCD.  This implies that rounding errors induced by the use of
finite precision in a digital computer may grow exponentially. Such
growth can be characterised in terms of the {\em leading Liapunov
exponent} of the system.  Furthermore, it has been shown
\cite{AdkHorvath} that the most commonly used MD integration scheme --
the leapfrog method -- has the potential to become unstable.
Instability is a problem for lattice QCD simulations since it results
in large energy changes along MD trajectories and hence negligible
acceptance rates in the HMC algorithm.

The instability in the leapfrog method has been illustrated in
\cite{AdkHorvath} for the case of free field theory where a mechanism
has been proposed which could explain the onset of such an instability
in lattice QCD. Numerical studies of the latter were carried out on
small lattices at a variety of couplings and quark masses.  The onset
of instability was found to be at smaller step sizes for lighter quark
masses.

Edwards, Horv\'ath and Kennedy \cite{AdkHorvath} also investigated an
optimisation strategy in which reduced work (and hence accuracy) in
the MD calculation was balanced against the resulting reduced
acceptance in the Metropolis step.  Each MD step requires the
iterative solution of a system of linear equations.  Since dynamical
fermion HMC codes spend a substantial fraction of their execution time
performing such solves it it clearly important to investigate whether
substantial efficiency gains can be made without introducing
undesirable effects such as loss of reversibility in the MD. The
investigation~\cite{AdkHorvath} was quite preliminary and the errors
quoted were quite large.  This issue was also investigated on small
lattices in \cite{ZbyshThesis}.  The present paper investigates many
of the issues raised in \cite{AdkHorvath} and extends the numerical
studies to production-scale lattices.

The paper is organised as follows.
In section \ref{s:HMC} we summarise the Hybrid Monte Carlo formalism 
and give details of the algorithms used.
Section \ref{s:Reversibility} contains
a discussion of the effects
of numerical roundoff errors on reversibility.  
In section \ref{s:Instability} we present 
results and discussion of our analysis of instability in the MD step.
In section \ref{s:TuningStudy2} 
we present the results of an optimisation analysis 
involving reduced accuracy in the MD step. 

Finally, in section \ref{s:Conclusions} we summarise our results and
conclusions.

\section{Hybrid Monte Carlo and lattice QCD}\label{s:HMC}
\subsection{HMC algorithm}
Consider a system with canonical coordinates $q$ and action $S(q)$.
One wishes to generate configurations $q$ with an 
equilibrium probability distribution in which
the statistical weight of configuration
$q$ is proportional to $e^{-S(q)}$.  

In Hybrid Monte Carlo, we introduce fictitious momenta $p$ conjugate
to $q$ and define a Hamiltonian function $H(q,p) = \frac{p^2}{2} +
S(q)$.

One may then generate configurations $(q,p)$ distributed according to
\begin{equation}
P(q,p) \ dq \ dp = \frac{1}{Z} e^{-H(q,p)} \ dq \ dp \qquad
\mbox{where} \qquad Z = \int dq \ dp \ e^{-H(q,p)} \ . 
\end{equation}
After the integration over the momenta, we obtain the desired distribution 
for the coordinates. 

Given an initial configuration $(q,p)$, a sequence of 
configurations is generated by repeated iteration of the following steps:
\begin{enumerate}
\item{{\bf Momentum refreshment: \ }}
Draw new fictitious momenta $p$ from a Gaussian distribution with
zero mean and unit variance.
\item{{\bf Molecular dynamics: \ }} 
Integrate the Hamiltonian equations of motion for some fictitious time
trajectory of length $\tau$, from the initial configuration $(q(0),p(0))=(q,p)$
to obtain the trial configuration $(q(\tau),p(\tau)) = (q',p')$.
\item{{\bf Accept/reject step: \ }}
The trial configuration $(q',p')$ is accepted with probability
\begin{equation}
P_{\rm acc}(q',p' \leftarrow q,p) = \min \left(1, e^{-\delta H}\right)
\end{equation}
where 
\begin{equation}
\delta H = H(q',p') - H(q,p) \ .
\end{equation}
If the trial configuration is rejected the new configuration is $(q,p)$.
\end{enumerate}

\subsection{Leap--frog integration}
For the HMC algorithm to satisfy detailed balance, the MD is required to
be reversible and measure preserving. This can be achieved through the
use of symmetric symplectic integration schemes, such as the
leapfrog algorithm.
In this algorithm, one constructs an approximation $\U_{3}(\dtau)$ to the time
evolution operator $\U(\dtau)$ for advancing a phase space vector $(q,p)$
through a step of length $\dtau$ in molecular dynamics time. 
The approximate operator $\U_{3}(\dtau)$ is itself 
composed of a symmetric combination
of the symplectic partial coordinate and 
momentum update operators $\U_{\rm q}(\dtau)$ and $\U_{\rm p}(\dtau)$ 
respectively, for example as
\begin{equation}
\U_{3}(\dtau) = \U_{\rm p} \left(\frac{\dtau}{2}\right) \U_{\rm q}(\dtau) \U_{\rm p}\left(\frac{\dtau}{2}\right) \ .
\end{equation}
The partial update operators are themselves defined as
\begin{eqnarray}
\U_{\rm q}(\dtau)\left(q,p\right) &=& (q + p \dtau, p) \\
\U_{\rm p}(\dtau)\left(q,p\right) &=& (q, p + F\dtau) 
\end{eqnarray}
where $F = -\frac{\partial S}{\partial q}$ is the MD force. Due to its
symmetric construction, $\U_{3}(\dtau)$ is reversible and, due to 
the symplectic nature of its component
updates, it is area preserving. 
The process of iteratively acting on an initial phase space vector with 
$\U_{3}(\dtau)$ is called leapfrog integration. The method is accurate
to $O(\dtau^3)$ per time step.

\subsection{Higher order integration schemes} 
\label{s:HOschemes}

The construction of higher order integration schemes (see 
for example \cite{Campostrini,Creutz})
is recursive, proceeding from the leapfrog
scheme. 
Given an approximate time evolution 
operator $U_{n+1}(\dtau)$ accurate to $O(\dtau^{n+1})$ 
for some even $n$, one can construct the 
operator 
\begin{equation}
\U_{n+3}(\dtau) = \U_{n+1}(\dtau_1)^{i}\U_{n+1}(\dtau_2)\U_{n+1}(\dtau_1)^{i}
\end{equation}
with 
\begin{eqnarray} 
\dtau_1 = \frac{\dtau}{2i - s} \label{e:dt1_eq} \\
\dtau_2 = \frac{\dtau}{1 - \frac{2i}{s}} \label{e:dt2_eq} 
\end{eqnarray}
where $i$ is an arbitrary positive integer and $s=(2i)^{1\over{n+2}}$. The step
sizes $\dtau_1$ and $\dtau_2$ are chosen to cancel truncation orders
of $O(\dtau^{n+1})$ and symmetry with respect to time 
ensures that there are no truncation
errors of $O(\dtau^{n+2})$. Hence such a scheme is accurate to $O(\dtau^{n+3})$.

Sexton and Weingarten~\cite{SextonWeingarten} have considered the
general case where the action $S$ can be split into two parts as $S(q)
= S_{1}(q) + S_{2}(q)$ and constructed an $O(\dtau^3)$ algorithm in
which the coefficient of leading order truncation error term may be
decreased.  The method is advantageous if evaluating the force
corresponding to $S_1$ is computationally much cheaper than the force
associated with $S_2$ (or vice versa). For example, one may take $S_1$
to be the gauge action and $S_2$ to be some computationally expensive
fermion action. The coefficient of the leading error term could then
be decreased by performing more gauge update steps than momentum
updates.

\subsection{Formulation of MD for lattice QCD}
The canonical coordinate variables for lattice QCD are the $SU(3)$
link matrices $U_{\mu}(x)$ associated with the link emanating from
site $x$ of the lattice and ending on neighbouring site $x+\hat{\mu}$,
where $\hat{\mu}$ is a unit vector in one of the Euclidean space--time
directions. The conjugate momentum fields $\pi_{\mu}(x)$ are members
of the Lie algebra $su(3)$.

In general, one can write the fictitious Hamiltonian for a lattice QCD system
with two degenerate flavours of Sheikholeslami-Wohlert (clover)
improved \cite{Clover,JansenSommer} fermions as
\begin{equation}
\tilde{H} = \frac{1}{2}\sum_{x,\mu} \pi_{\mu}^2  + S_{\rm g}(\beta; U) + \phi^{\dagger} \tilde{Q}^{-1}(\kappa, c; U) \phi \ ,
\end{equation}
where
\begin{equation}
\tilde{Q}(\kappa, c; U) = M^\dagger(\kappa,c;U) M(\kappa,c;U) \,.
\end{equation}
Here $M(\kappa,c;U)$ is the clover improved fermion matrix with improvement coefficient $c$, $\phi$ are pseudofermions  and  $S_{\rm g}(\beta; U)$ 
is the standard Wilson gauge action
\begin{equation}
S_{\rm g}(\beta; U) = - \frac{\beta}{6} \sum_{\Box}{\rm Re \ Tr \ } U_{\Box} \ . \label{e:WilsonGaugeAction}
\end{equation}
In (\ref{e:WilsonGaugeAction}) the sum is over all elementary
plaquettes $U_{\Box}$ on the lattice and $\beta = \frac{6}{g^2}$
where $g$ is the bare gauge coupling constant.

In our computations we have employed the technique of 
even--odd preconditioning which changes the form of $\tilde{Q}$
and $\tilde{H}$ somewhat. Each lattice site is labelled with a parity 
$p$ which is either {\em even} or {\em odd} so that any one lattice
site has an opposite parity from all of its neighbours. This allows
the fermion matrix to be block diagonalised and the Hamiltonian to be
re--written as:
\begin{equation}
H = \frac{1}{2}\sum_{x,\mu} \pi^{2} + S_{\rm g}(\beta; U) - 
2 \ \Tr \ln A_{e} + \phi^{\dagger}_{o} Q^{-1}(\kappa,c;U) \phi_{o}\, .
\end{equation}
Here, $A$ is the so called {\em clover} term summed over sites of one
parity (even in the equation above) and $Q$ is the {\em
preconditioned fermion matrix} coupling lattice sites of the opposite
parity (odd in the equation above) only. Thus $Q$ has half the rank of
$\tilde{Q}$. This leads to some memory saving at the additional expense of
having to evaluate $\Tr \ln A$ directly on sites of one 
parity. The precise formulation of the preconditioned matrices can be
found in \cite{StephsAndZbysh}.

We do not expect that splitting the Hamiltonian in this way will change
conclusions regarding reversibility and related issues in any
significant way. Although there is an extra force 
term to be computed to integrate
the equations of motion, the logarithm of the clover term is computed 
directly and is independent of the parameters used for the solution of the 
system of linear equations. Likewise,
for the inversion of the clover term, we use a direct method that is not
controlled by algorithmic parameters such as a target relative residue.
Hence we regard the effects of preconditioning as a minor technicality 
and shall disregard them for the rest of this paper.

The leapfrog partial update steps for the gauge fields and the momenta are
\begin{eqnarray}
\U_{q}(\dtau)\left( U_{\mu}(x), \ \pi_{\mu}(x) \right)  &=& 
\left( \exp \{ i \, \dtau \, \pi_{\mu}(x) \}  U_{\mu}(x), 
\, \pi_{\mu}(x) \right) \\
\U_{p}(\dtau)(U_{\mu}(x), \ \pi_{\mu}(x) ) &=& \left( U_{\mu}(x), 
\, \pi_{\mu}(x) + \dtau F_{\mu}(x) \right)
\end{eqnarray}
where 
\begin{equation}
F_{\mu}(x) = F^{\rm g}_{\mu}(x) + F^{\rm f}_\mu(x)
\end{equation}
and $F^{\rm g}$, $F^{\rm f}$ are the respective gauge
and fermionic force contributions,
\begin{eqnarray}
F^{\rm g}_{\mu}(x) &=& -\frac{\partial S_{\rm g}(U)}{\partial U_{\mu}(x)}  \\
F^{\rm f}_{\mu}(x) &=& \left[ Q^{-1} \phi \right]^{\dagger} 
\frac{\partial Q}{\partial U_{\mu}(x)} \left[ Q^{-1}\phi \right] \ .
\end{eqnarray}

\subsection{Solution of the linear system}\label{s:SolverIssues}
Computation of the fermion force 
requires the quantity
\begin{equation}
X = Q^{-1} \phi
\label{e:X} 
\end{equation}
which is obtained via the solution of the linear system
\begin{equation}
Q X = \phi \label{e:CGeqn}\, .
\end{equation}
This is normally carried out with a Krylov subspace solver such 
as the Conjugate Gradients (CG) \cite{CG} 
or the Stabilized BiConjugate Gradients (BiCGStab) \cite{BiCGStab} algorithm. 
With the BiCGStab solver, the solution consists of two solves:
\begin{eqnarray}
M^{\dagger}(\kappa,c) Y &=& \phi \label{e:BiCGStabYSolve}\\
M(\kappa,c) X &=& Y \label{e:BiCGStabXSolve}
\end{eqnarray}
whereas with CG, one can solve (\ref{e:CGeqn}) directly. When using
CG with a Hermitean positive definite matrix such as $Q$,
the solution is guaranteed to converge monotonically. With BiCGStab,
one has no such guarantee. Since the condition number of $Q$ is
the square of the condition numbers of either $M$ or $M^{\dagger}$,
we  expect the two stage solution using BiCGStab to be
faster on the whole than using one CG solve. As the convergence of 
BiCGStab can be erratic, it is prudent to restart the solution 
process for $X$ with CG using, as an initial guess, the solution for
$X$ from the previous BiCGStab solve. 

The solver residual $r_i$ at the $i$-th iteration of a CG solve
is defined as
\begin{equation}
r^{\rm Real}_i = \| \phi - Q X_{i} \| \label{e:RealResidue}
\end{equation}
where $X_{i}$ is the approximate solution at iteration $i$. The relative residual at the $i$-th iteration is then defined as
\begin{equation}
\rho^{\rm Real}_i = \frac{r^{\rm Real}_i}{\| \phi \|} \label{e:RealRelError} \ .
\end{equation}

In solver algorithms, $r_i$ is not usually computed using
(\ref{e:RealResidue}). Instead, $r_i$ is generally defined through
some three term or coupled two term recurrence relation.  We will
refer to this latter definition of the residual as $r^{\rm Acc}$,
the {\em accumulated residual}. The corresponding definition of the
relative residual is
\begin{equation}
\rho^{\rm Acc}_i = \frac{r^{\rm Acc}_i}{\| \phi \|} \label{e:CGRes} \ .
\end{equation}

These two definitions are equivalent in exact arithmetic.
However, computation of the accumulated residual needs only vector additions
and scalar multiplications whereas computation of the real residual
needs a matrix multiplication and so can differ in finite arithmetic.  
In our computations we use the
accumulated residual. We will denote by $r$ our target relative
residual. Hence the iterative process terminates when $\rho^{\rm
Acc}_i < r$. In the remainder of this paper we refer to $r$ as
the solver target residual, or just simply the solver residual.

\section{Reversibility}\label{s:Reversibility}
Reversibility and area preservation of the Molecular Dynamics step are
required for a correct HMC algorithm.
The leapfrog algorithm described in
section 2, is reversible and area preserving in exact
arithmetic. 
Computations are of necessity carried out
in finite precision and exact reversibility is lost. 
It is therefore important to verify that implementation of the
fundamental steps of the algorithm are as close to 
reversible as it is possible to make them.

Ideally, one would like to establish the least level of precision
required such that the accumulation of rounding errors does not
introduce a significant bias into the end results of a
calculation.  At present, it is not possible to give
a fully quantitative answer to this question. The accumulation of rounding
errors is a complex phenomenon and, since the underlying
equations of motion are known to be chaotic, the potential
for introducing large uncontrolled errors is great 
\cite{JansenLiu,AdkHorvath}. 
The best one can do is to ensure that
the implementation of each algorithmic component is
as close to reversible as practical and that the accumulation of errors grow 
in the expected way and so remain under control.

We study the reversibility of gauge and momentum 
update components separately.

\subsection{Gauge update}

The gauge update involves the process of exponentiating the conjugate
momenta on all lattice links~\cite{PieceOfPaper,Dubious}. 
One wishes to verify here that
\begin{itemize}
\item
the exponentiation of the momenta does produces a suitable 
unitary matrix;
\item
the exponentiation of the momenta is reversible in the sense
that
\begin{equation}
\exp(i \pi_{\mu}(x) \dtau) = \exp(-i\pi_{\mu}(x) \dtau)^{\dagger} \ .
\end{equation}
\end{itemize}

To check these properties, we studied
\begin{eqnarray}
\Delta {\rm Unit} &=& \max_{x,\mu,a,b}\left| \left(
\exp(i\pi_{\mu}(x)\dtau) \exp(i\pi_{\mu}(x)\dtau)^{\dagger} - 1
\right)_{ab} \right| \label{e:UnityViol}\\ \Delta {\rm Rev} &=&
\max_{x,\mu, a, b}\left| \left( \exp(i\pi_{\mu}(x) \dtau ) -
\exp(-i\pi_{\mu}(x)\dtau)^{\dagger}\right)_{ab} \right|
\label{e:HermViol} \ , \end{eqnarray} where $x$, $\mu$, $a$ and $b$ are
site, direction and colour indices respectively.  These observables
measure the maximum violations of unitarity and hermiticity on a
given lattice.

In tests of the gauge field update reversibility, 
we used quenched lattices with $V=4^4$ sites at $\beta=5.4$.
For the MD evolution we used $\tau = 1$ and $\dtau = \frac{1}{10}$.
The maximum values of both $\Delta {\rm Unit}$ and $\Delta {\rm Rev}$ 
along a molecular dynamics trajectory were found to be
\begin{equation}
\max_{\rm traj} \ \Delta {\rm Unit} = 
\max_{\rm traj} \Delta {\rm Rev} = 
0.59604635 \times 10^{-7} \approx \frac{1}{2}\epsilon_{SP} \ , 
\end{equation}
where $\epsilon_{SP}$ is the single precision unit of least precision.
The fact that the maxima of the metrics agree to 8 decimal places may seem
surprising at first, but becomes less mysterious when we recall
that we are working at the limits of single precision, where the
discrete nature of floating point numbers on a computer becomes apparent.
Hence, there is only a discrete set of numbers available that the metrics
can take of which the figure quoted above is one.

\subsection{Momentum update}
In the momentum update there are two possible sources of reversibility
violation. The first is a lack of associativity in the addition 
$p(\tau + \dtau) = p(\tau) + F(U)\dtau$
required in the update step. The second
arises in the computation of the force $F$. However, when performing a
momentum update forward in time for a step $\dtau$ followed
immediately by a momentum step backwards in time for $\dtau$, (with no
gauge field update in between) the gauge fields, and hence the force,
should remain unchanged. Thus, reversibility due to lack of associativity
in the addition can be isolated.

Consider a test where one starts with a set of fields $(U,\pi,\phi)$.
First the momentum fields are updated forward in time for a timestep $\dtau$
to produce fields $(U,\pi',\phi)$ and then a momentum update is performed
backwards in time\footnote{In practice this is done by flipping the signs
of all the momenta, integrating the equations of motion forward in time and flipping the signs of the momenta again.} to produce fields $(U,\pi'',\phi)$. We use the same value
of the force $F$ for both of the updates. One can then define the 
quantity
\begin{equation}
\Delta \pi^{i}_{\mu}(x) = \pi^{i}_{\mu}(x)'' - \pi^{i}_{\mu}(x)
\end{equation}
as a measure of the reversibility violation incurred by the momentum 
update step. To improve statistics, one may repeat this several times,
in each case using a new set of initial momenta drawn from a Gaussian 
distribution.

In the numerical tests, we started from some initial gauge field
configuration and performed MD in the ordinary sense. Before every
momentum update, we performed 100 forward--backward steps with
newly drawn momenta in each case. After the test was completed, we 
restored the original momenta from the end of the last gauge update step
and allowed the MD to continue. Thus we  obtained an estimate of
$\Expt{\Delta \pi^{i}_{\mu}(x)}$, the average reversibility
violation due to lack of associativity in the addition. At the end of
the complete trajectory, the resulting data was split into 8 sets,
one corresponding to each of the Lie algebra indices $i$.
The data in each set was histogrammed to obtain the distribution of the
average reversibility violation for each momentum component.

\begin{figure}[ht]
\begin{center}
\leavevmode
\hbox{%
\epsfxsize=4in
\epsffile{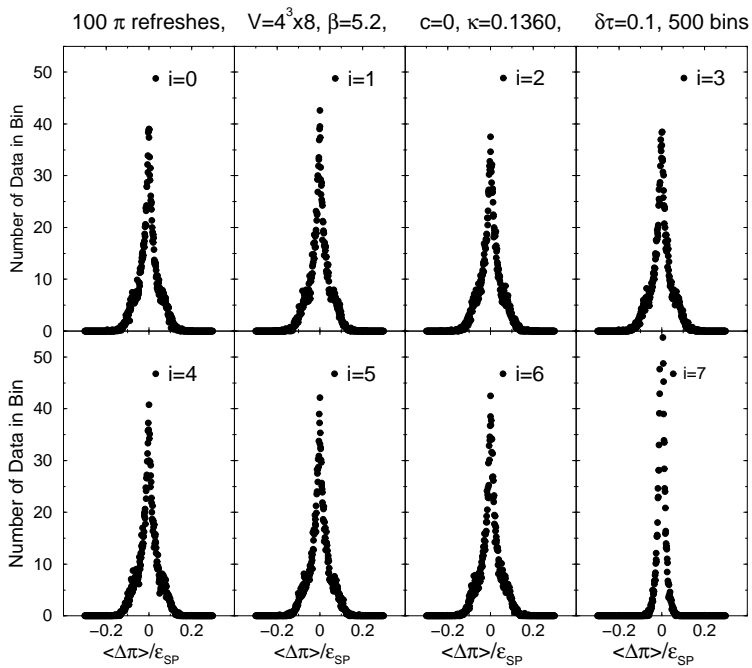}}
\end{center}
\caption{Distribution of momentum update reversibility violations obtained
by histogramming $\Expt{\Delta \pi}$. Each plot corresponds to a separate
momentum component and $\epsilon_{SP}$ is the single precision unit of least
precision.}
\label{f:MomChannels}
\end{figure}

The results of these momentum update tests are shown in 
figure \ref{f:MomChannels}. 
We show the histograms of all 8 momentum components. The errors
on the data points are small 
and, to aid clarity, 
are not displayed. 
The lattice volume used for these tests
was $V=4^{3}\times 8$ sites and
physical parameters were $\beta=5.2$, $c=0$ and $\kappa = 0.1360$.
We performed the tests following each gauge field
update along a trajectory consisting of 10 timesteps, each of length $\dtau=0.1$.
We used 500 bins for each momentum component in the histograms.
The histogramming process itself was carried out in double precision, 
allowing us to resolve reversibility violations of $O(10^{-1}\epsilon_{SP})$.

Figure \ref{f:MomChannels} shows that the distribution of
reversibility violations forms a very narrow, apparently symmetric,
distribution around $0$ with a width that is of
$O(10^{-1}\epsilon_{SP})$. We conclude that the momentum
update step in itself is as reversible as it is possible to
attain. The apparent symmetry of the distribution may possibly be used
to make more general statements about reversibility and area
preservation holding stochastically \cite{SextonStoch}.

\subsection{Reversibility of the force computation}
Since gauge fields are
unchanged along a momentum update and computation of the force due
to gauge fields is an entirely deterministic process, one 
expects that the force computation will be reversible. However, the
pseudofermion contribution to the force
requires solution of a linear equations, so further scrutiny is required.

It has been pointed out~\cite{Gupta} that the solution process should
be reversible, provided that a {\em time symmetric} initial guess
vector (such as a zero vector or vector with random components) is
used to start the solution process. This makes it tempting to carry
out such solves with a large target residue $r$, and hence save on the
computational workload. We discuss this further in
sections~\ref{s:SolverTuning} and \ref{s:TuningStudy2}.

Another commonly used solver strategy is to use 
the solution from the force computation of the previous momentum
update as an initial guess. This, and variants which use a more elaborate
extrapolation of previous solutions, may reduce the computational
workload but are inherently {\em non-reversible}
unless the solutions are effectively exact.

\subsection{Global Reversibility Violations}
Having discussed the sources of reversibility violation at a
microscopic level, we now turn to the problem of their global
accumulation. Consider an MD trajectory with initial
fields $(U,\pi)$ and a set of pseudofermion fields $\phi$. 
The latter remain unchanged along an MD trajectory. Suppose we perform
an MD trajectory forward to obtain fields $(U', \pi')$, then having
reversed the momenta, perform a second (backward) trajectory and a
momentum flip to obtain fields $(U'',\pi'')$. One may define the
following global reversibility violation metrics:
\begin{eqnarray}
\| \dDU \| &=& \sqrt{ \sum_{x,\mu,a,b} | U^{ab}_{\mu}(x)'' - U^{ab}_{\mu}(x) |^2 }  \ , \\
\| \dDP \| &=& \sqrt{ \sum_{x,\mu,i} \left(\pi^{i}_{\mu}(x)''-\pi^{i}_{\mu}(x) \right)^2 }  \ , \\
| \dDH | &=& \left| H(U'',\pi'',\phi) - H(U,\pi,\phi) \right| \ .
\end{eqnarray}
It is also useful to consider these quantities suitably normalised by their respective
degrees of freedom:
\begin{equation}
\|\dDU\|_{\rm d.o.f} = \frac{\| \dDU \|}{\sqrt{N^{U}_{\rm d.o.f}}} \ , \qquad 
\|\dDP\|_{\rm d.o.f} = \frac{\| \dDP \|}{\sqrt{N^{\pi}_{\rm d.o.f}}} \qquad \mbox{and} \qquad
| \dDH |_{\rm d.o.f} = \frac{| \dDH |}{\sqrt{N^{H}_{\rm d.o.f}}}\, .
\end{equation}
Here $N^{U}_{\rm d.o.f} = N^{\pi}_{\rm d.o.f} = 4 \times 8 \times V$ are the 
respective number of the gauge and momentum degrees of freedom (4 links per site
and 8 $SU(3)$ generators) and $N^{H}_{\rm d.o.f}$ is the number of degrees
of freedom involved in computing the Hamiltonian H. 
In the quenched approximation 
$N^{H}_{\rm d.o.f} = N^{U}_{\rm d.o.f} + N^{\pi}_{\rm d.o.f}$. When
dynamical fermions are included, there is an additional factor from the fermions
of $N^{f}_{\rm d.o.f} = 24 \times V$ (3 colour and 4 Dirac complex components
per site). In the even--odd preconditioned systems, half of the 
$N^{f}_{\rm d.o.f}$ degrees of freedom are represented in the
pseudofermion vectors and
the remainder absorbed into computing $\Tr \ln A$ on sites of the opposite
parity.

We also study $\frac{| \dDH |}{|\delta H|}$, where
\begin{equation}
\delta H = H(U', \pi') - H(U,\pi)\, .
\end{equation}
This is a measure of the relative error in our energy calculations
and is related to the accuracy of the acceptance probability. 
One would like this relative error to be quite small, certainly no more than 
a few percent.

\subsection{Volume scaling of global reversibility metrics}
According to their definitions, $\| \dDU
\|$ and $\| \dDP \|$ should scale as $O(\sqrt{V})$, since the metrics
require the summation of $O(V)$ positive definite quantities.
We therefore expect that the corresponding normalised (per degree of freedom)
metrics should volume-independent.
For $|\dDH|$, the summation involves numbers which are
not  positive-definite, and one might expect some cancellation.
If the numbers are truly
random, the cancellations between the terms can be modelled as a
random walk and one would expect the sum to scale as
$O(\sqrt{V})$.  Hence one would expect $|\dDH|_{\rm d.o.f}$ 
to be independent of the system volume in a manner similar to the 
$\| \dDU \|_{\rm d.o.f}$ and $\| \dDP \|$ metrics.

To satisfy ourselves further that our simulation code is performing as
well as can be expected, we carried out reversed trajectories (as described
in the definition of the metrics) in the quenched approximation with 
lattices of different volumes. In each case, we used a single configuration 
as the starting gauge field for the test and the momentum field was drawn
randomly from a heat bath. The trajectory length was $\tau = 1$
and the length of the timestep was $\dtau=\frac{1}{180}$. We used 
$\beta = 5.4$ and lattices of volume 
\begin{equation}
V \in \{ 4^4, 8^4, 10^3\times16, 16^3 \times 32 \} \ .
\end{equation}

\begin{figure}[ht]
\begin{center}
\leavevmode
\hbox{%
\epsfxsize=5in
\epsffile{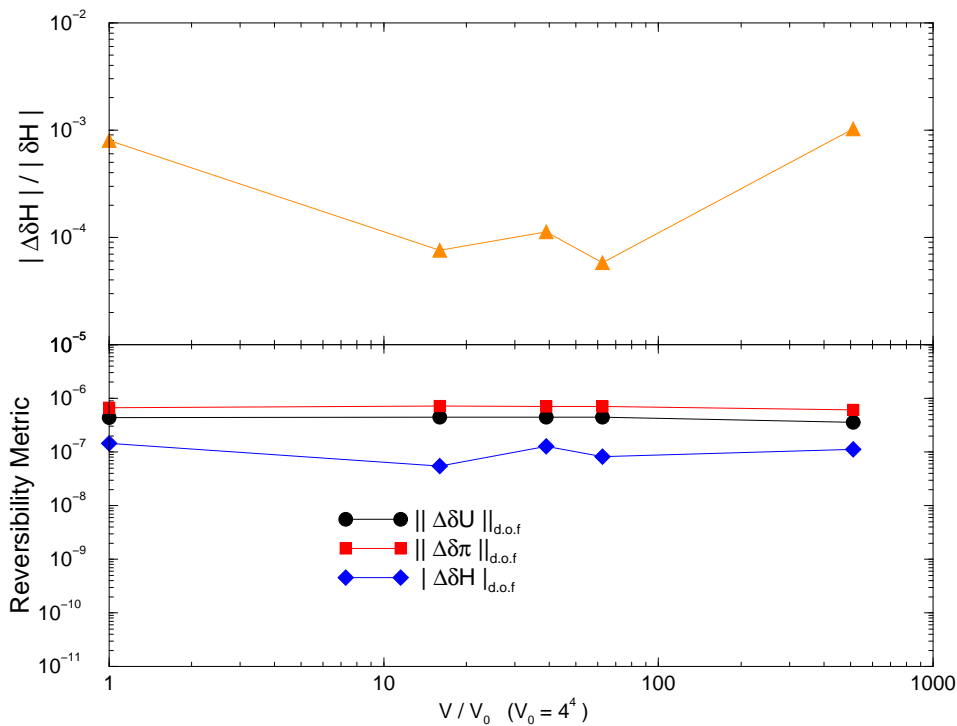}
}
\end{center}
\caption{Volume scaling of reversibility metrics, 
$\|\dDU\|_{\rm d.o.f}$, $\| \dDP \|_{\rm d.o.f}$, 
$| \dDH |_{\rm d.o.f}$ and $\frac{| \dDH |}{| \delta H |}$. 
The volumes are normalised by the smallest volume used:
$V_0=4^4$ sites.}
\label{f:VolumeScaling}
\end{figure}

Results of these tests are shown in figure \ref{f:VolumeScaling} where
the volumes have been normalised by the smallest one ($V_0 = 4^4$). 
We note that the degree of freedom normalised metrics -- $\| \dDU \|_{\rm d.o.f}$, $\|\dDP\|_{\rm d.o.f}$ and $\|\dDH\|_{\rm d.o.f}$ --
are all independent of the volume as expected. 
We also note 
that the relative error $\frac{|\dDH|}{|\delta H|}$ is less than of order
$0.1\%$, showing that error in computing the acceptance probability is
small.

\subsection{Accumulation of rounding errors in MD time}
It has been noted by several authors that
the MD equations of motion are chaotic \cite{JansenLiu,AdkHorvath}
and so effects of roundoff error are expected to grow exponentially 
with MD time along a trajectory. In particular, 
if one were to carry out reversed trajectory tests, as described in the
definition of the metrics $\| \dDU \|$ and $\| \dDP \|$, 
these would be expected to exhibit the leading behaviour
\begin{equation}
\|\dDU \| \propto e^{\nu_{U}\tau} \qquad \mbox{and} \qquad \| \dDP \| 
\propto e^{\nu_{\pi} \tau}
\end{equation}
as a function of the MD trajectory length $\tau$.
We use this as an operational definition
of the effective leading Liapunov exponents $\nu_{U}$ and $\nu_{\pi}$. In our
computations we measured only $\nu_{U}$ and, hence, in future discussion
we shall drop the subscript $U$ and refer to it simply as $\nu$. We shall
also refer to $\nu$ simply as the Liapunov exponent.

The authors of \cite{JansenLiu,AdkHorvath,ZbyshThesis} all found positive
values for the Liapunov exponents in their studies. In particular it was 
shown in \cite{AdkHorvath} that as the solver target residue $r$ and 
MD step--size $\dtau$ were made smaller, the Liapunov exponents
appeared to plateau, indicating that chaos was present in the underlying
continuum equations of motion for the system and not just a feature of
the numerical integration scheme. 

For the leading Liapunov exponent $\nu$, the authors of
\cite{AdkHorvath} found that this plateau came to an end at $\dtau
\approx 0.6$ in the quenched approximation and in the case of
dynamical fermion simulations with sufficiently heavy quarks. Beyond
this step size, the effective exponent exhibited growth.  However, in
the case of light quarks, this growth was found to set in
significantly earlier, at $\dtau\approx 0.08$.  This sudden growth in
Liapunov exponents could signal the onset of instability in the
MD. The subject of integrator instabilities will be taken up in
section~\ref{s:Instability}.

The authors of \cite{AdkHorvath} also studied the
behaviour of the Liapunov exponents as a function of the MD solver
target residue $r$. They
investigated the effects of increasing $r$ (using a time symmetric start) 
as a possible means of improving computational efficiency.
Their data
indicated a sudden growth in Liapunov exponent as 
$r$ is increased beyond a critical value. The data 
covered a limited range of $r$, and had large statistical errors.
However, the sudden apparent growth of the Liapunov exponent
coincides with a dramatic drop in acceptance rate, suggesting again that 
the integrator has become unstable.

\subsection{Tuning the solver target residual}\label{s:SolverTuning}
The results of \cite{AdkHorvath} motivated us to measure the Liapunov
exponents of our simulations while varying the target residue of a comparatively
large volume system, with comparatively light quarks such as those
in current production runs. 

For the determination of Liapunov exponents, we used 10
configurations taken from one of our large data sets. The lattice volume used
was $V=16^3 \times 32$ and the physical parameters were $\beta=5.2$,
$c=2.0171$ and $\kappa=0.1355$.  The value of the clover coefficient
was calculated using the formula determined by the Alpha collaboration
\cite{JansenSommer}.  These parameters correspond to pseudoscalar
to vector mass ratio of $\frac{m_\pi}{m_\rho}\approx 0.6$
\cite{NewUKQCD} and a lattice spacing of $a=0.097\fm $~\cite{NewUKQCD}
where the physical lattice spacing has been determined using the
observable $r_0$\cite{Sommer}. By current standards, 
the dynamical fermions are relatively light.

Using the 10 starting configurations, for a given value of $r$ we
carried out reversed MD trajectories of varying length $\tau$ with a
constant step--size of $\dtau = \frac{1}{180}$.  This value for
$\dtau$ was the one used in the production of the dataset from which
our 10 sample configurations were taken. Our MD 
solver strategy was to employ a two stage BiCGStab solution to compute
the quantity $X$ of (\ref{e:X})  
followed by a restarted CG solution. Hence the target
residue used was the accumulated target residue for the 
CG solver as described in section \ref{s:SolverIssues}. 
The target residues used ranged from $r=10^{-7}$ to $r=10^{-4}$. The smallest
of these is near the limit of what may be achieved in a single precision
(32bit) computation.

In each test we measured $\|\dDU\|$, $|\delta H|$ and $N_{\rm iters}$, 
where $N_{\rm iters}$ was the total number of solver iterations carried
out in both the BiCGStab and CG solves averaged over the forward and reverse
trajectories. For each combination of parameters, we also calculated the 
Metropolis acceptance probability $P_{\rm acc}$. 

To evaluate the savings (or losses) in computational cost we defined
the cost metric
\begin{equation}
\mbox{Cost} = \frac{N_{\rm iters}}{P_{\rm acc}} \ \label{e:cost_Func} .
\end{equation}
This heuristic measure reflects the fact that a large number of iterations
along an MD trajectory implies high computational cost, as does a low 
Metropolis acceptance rate. We note that an absolute measure of cost should
also take into account the autocorrelation time of the ensemble produced
by an HMC computation. Since we are unable to control or measure this 
quantity on a sample of 10 configurations, we disregard autocorrelation
effects in this study where we are interested in the {\em relative}
cost with different choices of simulation parameters.

\begin{figure}[ht]
\begin{center}
\leavevmode
\hbox{%
\epsfxsize=4.5in
\epsffile{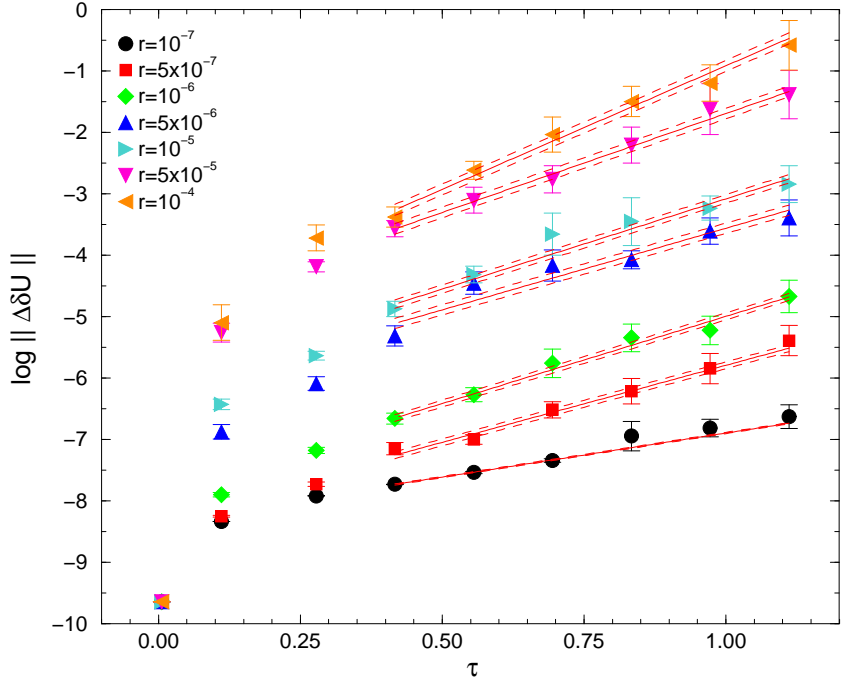}}
\end{center}
\caption{Fits for the Liapunov Exponent $\nu$.}
\label{f:LiapunovFits}
\end{figure}

Figure \ref{f:LiapunovFits} shows fits used to extract the (effective) 
Liapunov exponents. The system is clearly chaotic as $\ln \|\dDU\|$ 
has a significant positive slope as a function of $\tau$. 
Even with only 10 configurations, the signal for the Liapunov exponents is
good except for the cases when $r=5\times10^{-6}$ and when $r=10^{-5}$. 
The data for these latter parameter values seem to show a marked break 
at $\tau \approx 0.6$ and indeed, it was not possible to establish a 
consistent value of the Liapunov exponent for these two values of $r$.

\begin{figure}[ht]
\begin{center}
\leavevmode
\hbox{%
\epsfxsize=4.5in
\epsffile{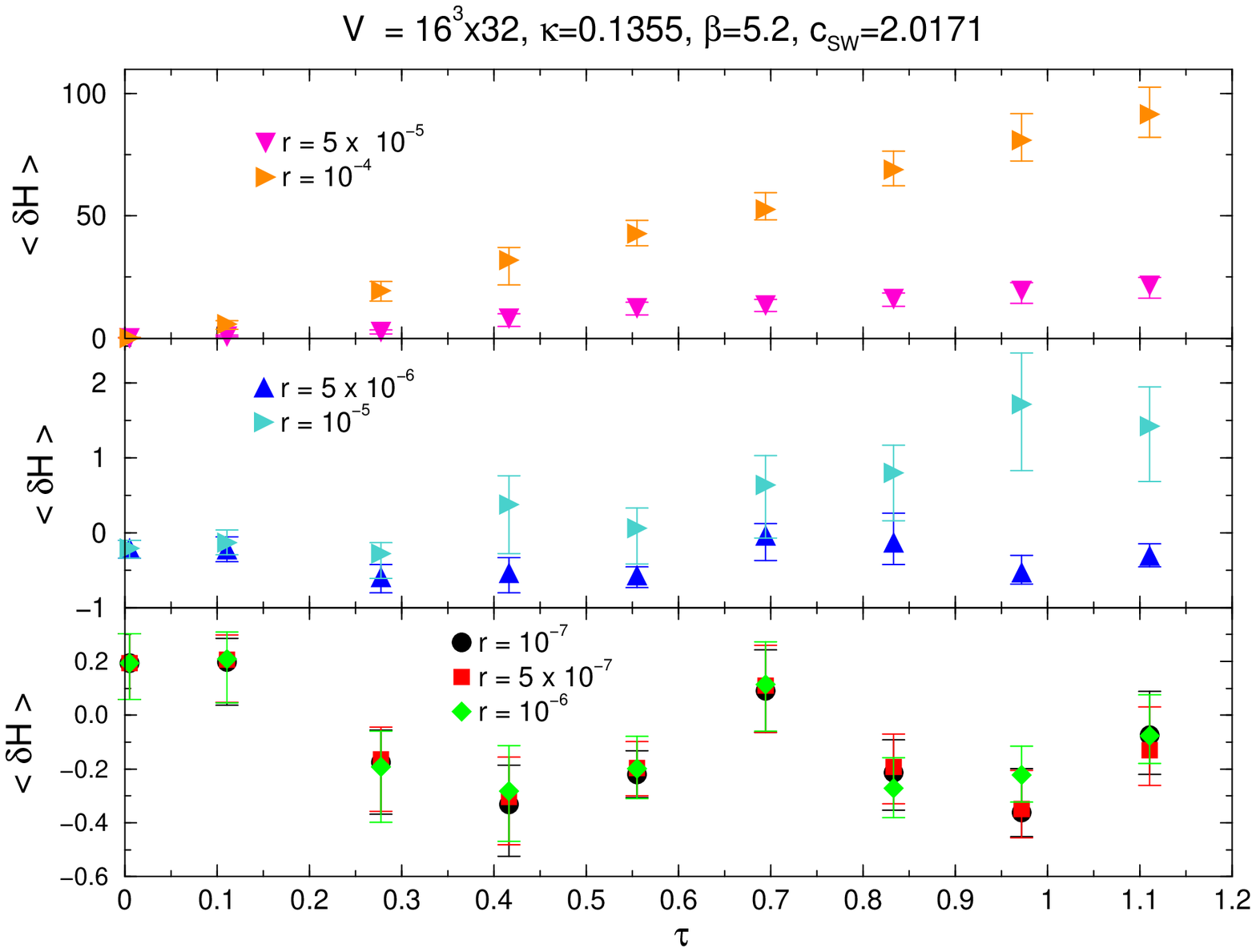}}
\end{center}
\caption{$\Expt{\delta H}$ as a function of $\dtau$ for various values of $r$.}
\label{f:H_vs_tau}
\end{figure}

In figure \ref{f:H_vs_tau} we show $\Expt{\delta H}$, the energy change 
along an MD trajectory
averaged over 10 configurations as a function of trajectory length $\tau$.
One can clearly distinguish three different types of behaviour 
for $\Expt{\delta H}$
depending on the target MD residual $r$. 
For values of $r < 5\times 10^{-6}$, $\Expt{\delta H}$ shows an 
oscillatory behaviour with $\tau$, 
whereas for $r > 10^{-5}$ $\Expt{\delta H}$
diverges with increasing $\tau$, resulting in 
a corresponding exponential drop in 
acceptance probability. 
It is interesting to note that this change in the
behaviour of $\delta H$ occurs at the value of $r$ where the data
in figure \ref{f:LiapunovFits} also show a change.

\begin{figure}[ht]
\begin{center}
\leavevmode
\hbox{%
\epsfxsize=4.5in
\epsffile{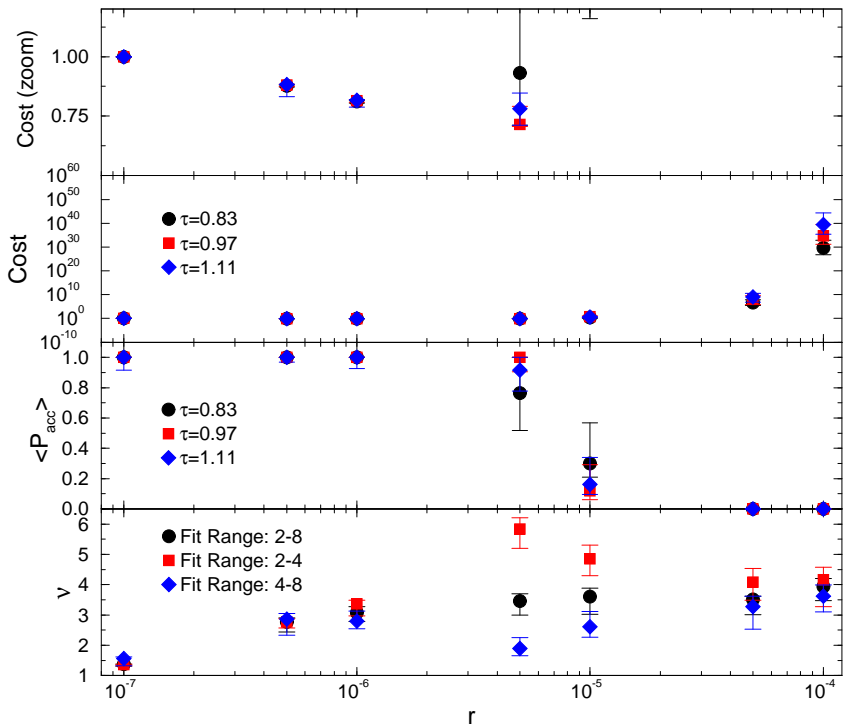}}
\end{center}
\caption{Liapunov exponents, acceptance rate and cost as a function of $r$.}
\label{f:all_vs_r}
\end{figure}

A summary of results for tuning the solver
residue is shown in figure \ref{f:all_vs_r}. 
The bottom panel
shows the Liapunov exponents $\nu$. For each value of $r$ we made
several determinations of $\nu$ by fitting to different ranges of
$\tau$ in figure \ref{f:LiapunovFits}. We note that the results of
these different fits are consistent with each other except for the
values of $r = 5\times 10^{-6}$ and $r = 10^{-5}$ corresponding to the
``break'' evident in figure \ref{f:LiapunovFits}. 

We note that, overall, the Liapunov exponents appear to show a slow
growth with $r$. There is no evidence of a plateau as $r$ is reduced to
$r=10^{-7}$.  This implies that this manifestation
of chaos in the system is {\em not} due
to the underlying equations of motion, but to the
integrator. The behaviour of the exponents near $r=10^{-5}$ may
perhaps be interpreted as the effect of the integrator changing
from being stable to being unstable.

The second panel in figure \ref{f:all_vs_r} shows the
average acceptance rate $\Expt{P_{\rm acc}}$ for
trajectories of length $\tau \approx 1$.  The acceptance 
shows a rapid drop for $r > 10^{-5}$, which is
due to the divergent behaviour of $\delta H$ for values of $r$
in this region.  The rapid drop in acceptance rate results in a
huge growth in the cost of the algorithm as shown in the third panel
of figure \ref{f:all_vs_r} where we display the cost metric~(\ref{e:cost_Func})
normalised by its value for the simulation with $r=10^{-7}$.

In the top panel of figure \ref{f:all_vs_r} we show an enlarged view
of the cost function for values of $r < 10^{-5}$. 
The cost metrics for values
of $r \ge 10^{-5}$ are too large to fit onto this enlarged plot. We note
that the normalised cost has a shallow 
minimum when $r=5 \times 10^{-6}$ however at this
minimum value the normalised cost has a value of about 0.75 implying a saving
of only about $25\%$. 

\section{Instability in the MD integration}\label{s:Instability}
The behaviour of the energy change $\delta H$, from oscillatory to
divergent, is reminiscent of a known instability in the
leapfrog algorithm when applied to the integration of the equations
of motion for the simple harmonic oscillator. In this section, we review
the simple harmonic oscillator analysis of~\cite{AdkHorvath} and 
compare expectations for interacting theories with our numerical results.
\subsection{Harmonic Oscillator}
In what follows we use the
notation of ~\cite{AdkHorvath}.
Consider a single 
oscillator with coordinate $\phi $. The corresponding Hamiltonian function is
\begin{equation}
H = {1 \over 2} \left( \pi^2 + \omega^2 \phi^2 \right) \ , 
\end{equation}
where $\omega$ is the angular frequency of the oscillator and
$\pi$ is the corresponding fictitious momentum.

The leapfrog update for the coordinate and momentum may be
written in the form of a matrix ${\mathcal{U}}_{3}(\delta \tau)$ acting
on the phase space vector $(\phi, \pi)$
\begin{equation}
{\mathcal{U}}_{3}(\delta \tau) = \left( \begin{array}{cc} 
1 - \frac{1}{2}\omega^2 \delta \tau^2 & \delta\tau \\ 
-\omega^2 \delta \tau + \frac{1}{4}\omega^4 \delta 
\tau^3 & 1 - \frac{1}{2} \omega^2 \delta \tau^2 \end{array} 
\right)\, . \label{e:Ueqn}
\end{equation}

The update matrix $\U_3$ can be parameterised as
\begin{equation}
{\mathcal{U}}_{3}(\delta \tau) = \left( \begin{array}{cc} 
\cos[\kappa(\delta \tau)\delta \tau] & \displaystyle 
\frac{\sin[ \kappa( \delta\tau)\delta \tau]}{\rho( \delta \tau)} \\  
-\rho( \delta \tau)\sin[\kappa( \delta \tau)\delta \tau] & 
\cos[\kappa( \delta \tau)\delta\tau] \end{array} \right)
\end{equation}
where
\begin{eqnarray}
\kappa(\delta \tau) &=& 
\frac{\cos^{-1}( 1 - \frac{1}{2}\omega^2 \delta \tau^2)}{\delta \tau} 
\label{e:kappadef} \\
\rho(\delta \tau)  &=& \omega\sqrt{1 - 
\frac{1}{4}\omega^2 \delta \tau^2} \label{e:rhodef} \ .
\end{eqnarray}
Evolution over a whole trajectory of length $\tau$ is then given by
\begin{equation}
{\mathcal{U}}_{3}(\tau) = \left( \begin{array}{cc} \cos[\kappa( \delta
\tau)\tau] & \displaystyle \frac{\sin[ \kappa( \delta\tau) 
\tau]}{\rho( \delta \tau)} \\  
-\rho( \delta \tau)\sin[\kappa( \delta \tau)\tau] & 
\cos[\kappa( \delta \tau)\tau] \end{array} \right)
\label{e:EvolutionParameterisation} \ .
\end{equation}

The nature of the instability in the leapfrog
scheme may be illustrated by examining the phase space trajectories
in this system.
The initial phase space vector 
for an oscillator released from amplitude $A$ is
$(\phi(0), \pi(0)) = (A,0)$. From (\ref{e:EvolutionParameterisation}),
the phase space vector at time $\tau$ is then
given by
\begin{equation}
\left( \begin{array}{c} \phi(\tau) \\ \pi(\tau) \end{array} \right) =
\left( \begin{array}{c} A\cos[ \kappa( \delta \tau) \tau] \\ 
- A\rho( \delta \tau) \sin [ \kappa( \delta \tau) \tau]  
\end{array} \right)\, .
\end{equation}
The phase space orbits therefore satisfy
\begin{equation}
\frac{\phi^2(\tau)}{A^2} + 
\frac{\pi^2(\tau)}{A^2\rho^2(\delta \tau)} = 1 \, .
\label{e:ConicTraj}
\end{equation}

It can then be seen from 
(\ref{e:rhodef}) and 
(\ref{e:ConicTraj}) 
that for $\omega \delta \tau < 2$ 
the phase space trajectories are elliptical\footnote{In the exact solution
the orbits are circular, the deformation to an ellipse is an effect of the
truncation error in the leapfrog scheme even in exact arithmetic.}, 
whereas for $\omega\delta \tau > 2$
they are hyperbolic. The instability at $\omega \delta \tau = 2$ is the abrupt 
transition from one class of phase space trajectories to another.

The change in energy
\begin{equation}
\delta H = H(\phi(\tau), \pi(\tau)) - H(\phi(0),\pi(0)) 
\end{equation}
may also be computed.
Using the same initial conditions,
\begin{equation}
\delta H = -\frac{1}{8}\omega^4 A^2 \dtau^2 
\sin^2[\kappa(\delta\tau)\tau]\, .
\label{e:FreeFieldDH}
\end{equation}
When $\omega \delta \tau < 2$, $\kappa(\dtau)$ is real and so $\delta
H$ oscillates with increasing $\tau$, in a manner similar to that
observed in the bottom panel of figure \ref{f:H_vs_tau}.  However,
when $\omega \dtau > 2$, $\kappa(\dtau)$ becomes purely imaginary
causing $\delta H$ to diverge as ${\rm
sinh}^2[\kappa(\delta\tau)\tau]$ in a manner similar to that seen in
the top panel of figure~\ref{f:H_vs_tau}.

\subsection{Generalised treatment of instabilities}
We now present a more
general method of finding instabilities in the leapfrog algorithm
and in higher order schemes of the type discussed in \cite{Campostrini,Creutz}
(see section~\ref{s:HOschemes}) 
when applied to the case of a harmonic oscillator.

Consider an initial phase space vector $(\phi, \pi)$ of the harmonic
oscillator. This is to be evolved through phase space by the
leapfrog matrix ${\mathcal{U}}_{3}(\dtau)$ of (\ref{e:Ueqn}).  The
area preservation property of the integrator implies that
$\det({\mathcal{U}}_{3}(\dtau)) = 1$. All components of
${\mathcal{U}}_{3}(\dtau)$ are real, implying that $\Tr \
{\mathcal{U}}_{3}$ is also real.

If
\begin{equation}
\lambda_{1} = u_1 + i v_1 \qquad \mbox{and} \qquad \lambda_{2} = u_2 +
i v_2 
\end{equation}
are the two eigenvalues of ${\mathcal{U}}_{3}(\dtau)$,
the previous conditions on the trace and the determinant (area preservation) 
can then 
be shown to imply that
\begin{equation}
v_1 = -v_2 \quad\hbox{and}\quad u_1 v_2 + u_2 v_1 = 0\ . \label{e:conds}
\end{equation}
We conclude that either:
\begin{enumerate}
\item[1)]{$u_1 = u_2$ \qquad\hbox{or}\ } 
\item[2)]{$v_1 = v_2 = 0$ \ } 
\end{enumerate}
	In case 1), the determinant condition ($\lambda_1\lambda_2=1$)
implies that $u_1^2 + v_1^2 = 1$.  The eigenvalues have magnitude
unity: $\lambda_{1,2} = e^{\pm i\theta}$ with $\theta$ real, and the
update matrices ${\mathcal{U}}_{3}(\dtau)$ and
${\mathcal{U}}_{3}(\tau)$ ($= {\mathcal{U}}_{3}^{N_{\rm MD}}(\dtau)$)
give stable elliptical trajectories in phase space.


In case 2), by the same condition on the determinant, we have that
$\lambda_1 = \eta$ and $\lambda_2 = \frac{1}{\eta}$ for some real
$\eta\ge1$. On raising $\lambda_1$ or $\lambda_2$ to the power $N_{\rm
MD}$, one of the eigenvalues of ${\mathcal{U}}_{3}(\tau)$ will show an
exponential divergence with $N_{\rm MD}$. This implies unstable
behaviour in the integrator.

The condition for the onset of instability is that the eigenvalues
change from being complex to real. This information can be deduced
from the discriminant of the characteristic polynomial of the update
matrix ${\mathcal{U}}_{3}(\dtau)$. The onset of instability occurs as
the discriminant changes sign from negative to positive.

For the leapfrog method, the discriminant 
is given by
\begin{equation}
D_3 = ( \omega \dtau )^2( \omega \dtau - 2)(\omega \dtau + 2)\, .
\end{equation}
We note that for 
$0 < \omega \dtau < 2$, the discriminant is negative indicating 
a stable integrator, whereas for $\omega \dtau > 2$ the discriminant 
is positive implying an unstable integrator in line with the previous
discussion.

\subsection{Instability in Higher Order Schemes}
Consider the 5th order scheme of Campostrini and Rossi~\cite{Campostrini}.
This can be constructed from three leapfrog integration steps as
\begin{equation}
{\mathcal{U}}_{5}(\dtau) = 
{\mathcal{U}}_{3}(\dtau_1){\mathcal{U}}_{3}(\dtau_2){\mathcal{U}}_{3}(\dtau_1)
\end{equation} 
with $\dtau_1 = \frac{\dtau}{2-2^{1/3}}$ and $\dtau_2 = 
-\frac{2^{1/3}\dtau}{2 - 2^{1/3}}$. This corresponds to 
$n=3$ and $i=1$ in (\ref{e:dt1_eq}) and (\ref{e:dt2_eq}).

\begin{figure}[ht]
\begin{center}
\leavevmode
\hbox{%
\epsfxsize=4.5in
\epsffile{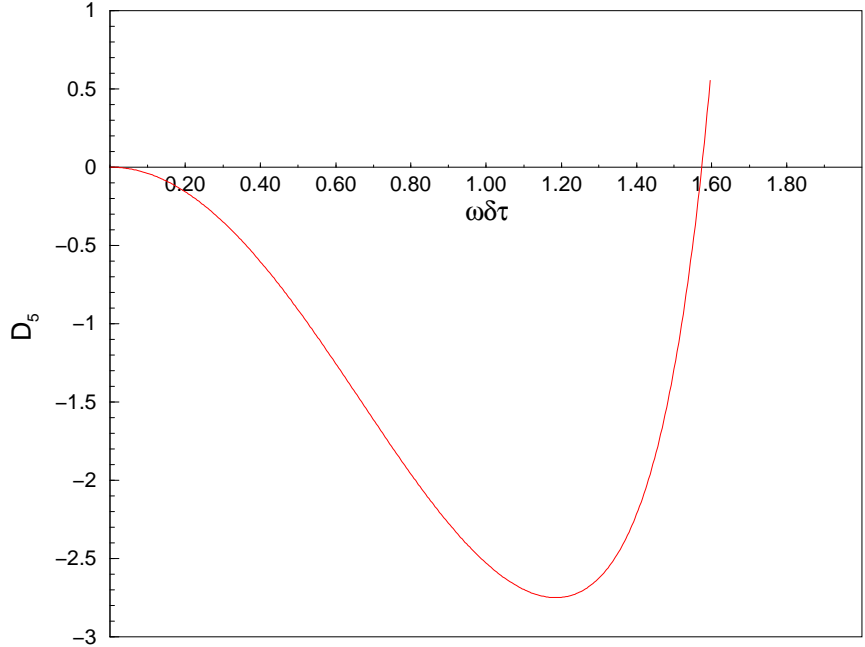}}
\end{center}
\caption{The discriminant $D_5$ of characteristic 
polynomial of the 5th-order Campostrini-Rossi update matrix ${\mathcal{U}}_{5}(\dtau)$.}
\label{f:CampostriniDiscrim5}
\end{figure}

The discriminant $D_5$ is a 12th order polynomial in $\omega \dtau$
which can easily be found using an algebraic package such as Maple.
It is not reproduced here but plotted in figure \ref{f:CampostriniDiscrim5}.
The nonnegative roots of the $D_5 = 0$ are found to be
\begin{equation}
\omega \dtau \in \{ 0, \sqrt{12-6\sqrt[3]{4}}\}\, .
\end{equation}
To three decimal places, the positive root is at $1.573$.
The discriminant is negative for $0 < \omega \dtau < 1.573$ 
indicating stable behaviour and is positive for $\omega \dtau > 1.573$ 
for the region where the integrator is unstable. 

It is interesting to note that, for  the central leapfrog
update matrix ${\mathcal{U}}_{3}(\dtau_2)$ in the 5th order scheme to
become unstable on its own, requires that $\omega \dtau_2 = 2$. This
implies that this central step should go unstable when 
\begin{equation}
\omega \dtau = 2 \frac{\left( 2 - 2^{1/3} \right)}{2^{1/3}} \approx 1.175 \ .
\end{equation}
This suggests that, although the central update itself becomes
unstable at $\dtau = 1.175$, the other two updates in the scheme
stabilize the system until $\dtau \approx 1.57$. 

Following a similar calculation, it can be shown that
the discriminant $D_7$ of the characteristic polynomial for the update
matrix of the 7th order scheme ($n=5$, $i=1$) has roots at
\begin{equation}
\omega \dtau \in \{ 0, 1.595, 1.822, 1.869 \}
\end{equation}
with $D_7$ being negative in the intervals $D_7 \in (0,1.595)$
and $D_7 \in (1.822,1.869)$ indicating two domains of stability.
The discriminant is positive for $D_7 \in (1.592,1.822)$ and 
for $D_7 > 1.869$. For the longest constituent 5th order update to
go unstable in this scheme requires that $\omega \dtau > 1.166$.

Hence we see that, for the case of the simple harmonic oscillator
at least, higher order integration schemes do not help cure the problem
of instabilities. Indeed, they become unstable at even smaller values
of $\omega \dtau$ than the simplest leapfrog method.

\subsection{Hypothesis for interacting field theories}
Edwards, Horv\'ath and Kennedy~\cite{AdkHorvath} 
advanced the hypothesis 
that, since the high frequency modes of an
asymptotically free field theory can be considered as a collection of
weakly coupled oscillator modes, the instability just
described in the harmonic oscillator system will also be present
for interacting field theories.
The onset of the instability will be caused by the mode with highest
frequency $\omega_{\rm max}$, when $\omega_{\rm max} \delta \tau =
2$. For a single oscillator mode, the onset of instability is
abrupt. In the case of an interacting theory, one would expect the
effects of the interactions to smooth out this transition.

It is argued in~\cite{AdkHorvath} that the instability in 
lattice QCD with dynamical
fermions can be likened to that of a collection of oscillator modes of the
sort just described. When applying leapfrog integration to this
system, the r\^ole of $\omega^2\phi$ in the harmonic oscillator example is
played by the MD force $F_{\mu}(x)$. This force can be
written as a sum of contributions from the gauge and fermionic 
pieces of the action as $F_{\mu}(x) = F^{\rm g}_{\mu}(x) + F^{\rm
f}_{\mu}(x)$, where the subscripts $g$ and $f$ indicate the
gauge and fermionic components of the force respectively.

The fermion force is expected to be proportional to $m_f^\alpha$,
where $m_f$ is the mass of the lightest species of dynamical fermion
and $\alpha$ is some \emph{negative} parameter.
In the case of Wilson (and Clover) fermions the mass in lattice units
is defined as
\begin{equation}
am_f = \frac{1}{2}\left( \frac{1}{\kappa} - \frac{1}{\kappa_c} \right) \label{e:WilsonMass}
\end{equation}
where $\kappa$ now stands for the Wilson hopping parameter, and
$\kappa_c$ is the critical value corresponding to $m_f = 0$. 
It is argued 
that the highest frequency mode (with frequency $\omega_{\rm max}$) is
proportional to the fermion force which, in turn, is expected to be proportional
to $m_f^{\alpha}$, and thus as $\kappa \rightarrow \kappa_c$ ($m_f
\rightarrow 0$), the fermion force will diverge and hence the critical
value of $\delta \tau$ will decrease. 
In the following, we evaluate numerical evidence 
for the validity of this hypothesis.

\subsection{Studies of the force}
The forces used in the momentum update belong to the
Lie algebra $su(3)$.
We define the 2--norm $\|F\|$ in the
same manner as for $\| \dDP \|$:
\begin{equation}
\| F \| = \sqrt{ \sum_{x,\mu,i} \left( F^{i}_{\mu}(x) \right)^2 } \ .
\end{equation}
Again, we can define the 2--norm
suitably normalised by the relevant degrees of freedom:
\begin{equation}
\| F_{g} \|_{\rm d.o.f} = \frac{\| F_{g} \|}{\sqrt{N^{U}_{\rm d.o.f}}}
\qquad \mbox{and} \qquad \| F_{f} \|_{\rm d.o.f} 
= \frac{\| F_{f} \|}{\sqrt{N^{f}_{\rm d.o.f}}}
\end{equation}
where the subscripts $g$ and $f$ indicate
gauge and fermionic forces respectively.

We can also define an $\infty$--norm for the forces:
\begin{equation}
\| F \|_{\infty} = \max_{x,\mu,i} \left| F^{i}_{\mu}(x) \right| \ .
\end{equation}

The $\infty$--norm then is the force component with the maximum
magnitude over the lattice and so can be likened to the force mode
with the highest frequency, proportional to $\omega^2_{\rm max}$, in
the analogous collection of weakly coupled harmonic oscillators.  The
(degree of freedom) averaged 2--norm on the other hand can be likened
to the average frequency--squared of the analogous set of harmonic
oscillators.

In our studies we computed the magnitude of the forces at
all timesteps of an MD trajectory starting from a single gauge
configuration chosen from the same 10 configurations 
described in section \ref{s:SolverTuning}
(with volume lattice $V=16^3\times32$ sites, and production parameters
$\beta=5.2$, $c=2.0171$, $\kappa=0.1355$). 

In the first set of tests, we attempted to investigate how the fermion
force behaves with the quark mass. 
We performed MD
trajectories consisting of $N_{MD}=175$ steps of length $\dtau =
\frac{1}{180}$ for several values of the hopping parameter
$\kappa$. We measured the norms of the gauge and fermion forces on
each timestep. The MD solver target residue was set at $r =
10^{-6}$. Error bars for the average value of the force were computed
by bootstrapping the 175 samples.

It could be argued that a configuration that has been 
produced in an ensemble equilibrated at some value of $\kappa$, 
will have very small statistical weight at a different value of $\kappa$.
However, our aim was not to study equilibrium properties of the
ensemble, but to test the properties of 
algorithm components  as a function of the external parameter $\kappa$.

The average value of $\kappa_{c}$, the critical value of $\kappa$
corresponding to massless fermions, is known from separate
spectroscopy studies for the ensemble from which the configurations
were drawn. It is approximately $0.1363$ \cite{NewUKQCD}. 
Thus, we were able
to associate a value of the lattice fermion mass $a m_{f}$ with every value 
of $\kappa$ used in our tests through the formula:
\begin{equation}
am_f = \frac{1}{2}\left( \frac{1}{\kappa} - \frac{1}{\kappa_c} \right) \ .
\end{equation}

Since we expect the fermion mass to vary in 
some inverse relation to the norm of the force~\cite{AdkHorvath},
we attempted to fit the 
results of our tests with the form 
\begin{equation}
F = A ( a m_{f} )^{\alpha} = A \left(\frac{1}{2\kappa} - 
\frac{1}{2\kappa_c}\right)^{\alpha} \ , \label{e:FitAnsatz}
\end{equation}
where the parameters of the fit were $A$, $\kappa_{c}$ and $\alpha$.

\begin{figure}[ht]
\begin{center}
\leavevmode
\hbox{
\epsfxsize=4.5in
\epsffile{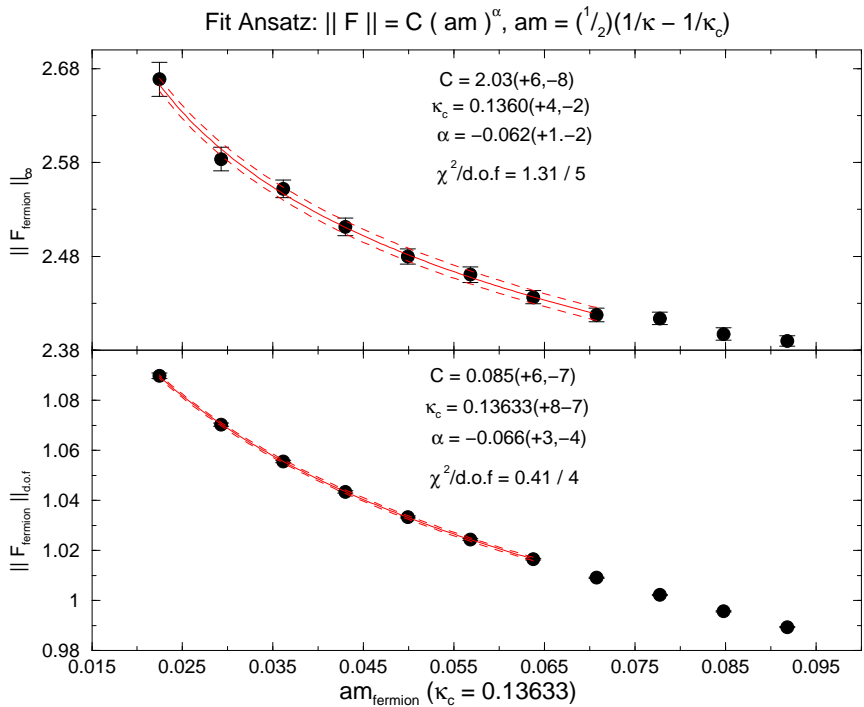}
}
\end{center}
\caption{Fits to the fermion force as a function of $am_f$ using the 
fitting hypothesis of (\ref{e:FitAnsatz}).}
\label{f:ForceFits}
\end{figure}

Results of this test are shown in figure \ref{f:ForceFits}. We show
both the fits made to the $\infty$--norm and the (degree of freedom) averaged
2--norm of the force. We can see that good fits can be made, which reproduce 
$\kappa_c$ from the spectroscopic studies and that $\alpha$ is negative
indicating that the magnitudes of the norms do indeed vary in an 
inverse manner with the fermion mass.
The fact that the value of $\kappa_c$ is well reproduced
and that $\alpha$ is negative in 
sign both lend support to the hypothesis of \cite{AdkHorvath}.

\subsection{Dependence on $\dtau$ and $\kappa$}
\begin{figure}[ht]
\begin{center}
\leavevmode
\hbox{%
\epsfxsize=4.5in
\epsffile{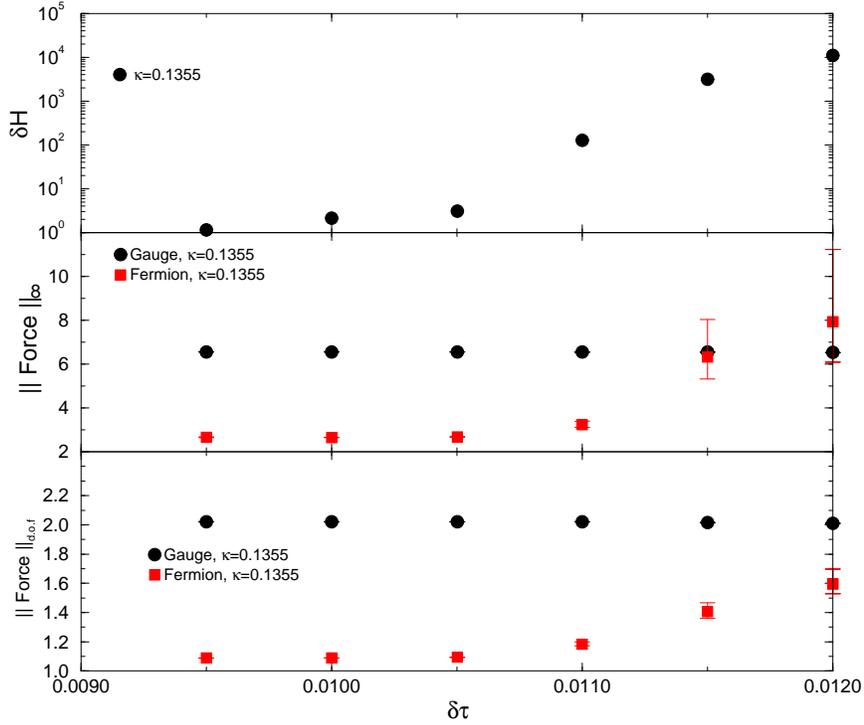}}
\end{center}
\caption{The 2--norm and the $\infty$--norm of the average gauge and
fermionic components of the MD force along an MD trajectory plotted
against the MD stepsize $\dtau$. The corresponding behaviour of the
energy change $\delta H$ along a trajectory is shown in the top graph.}
\label{f:Forces_vs_dtau}
\end{figure}

To investigate further the onset of instability, we computed the 
averaged forces and $\delta H$ along an MD trajectory using the same
starting configurations as before.  
However, this time we varied the MD
step size $\delta \tau$. The number of steps taken along the trajectory
was adjusted to keep the trajectory length constant at $\tau =175/180$.
The results are plotted in figure \ref{f:Forces_vs_dtau}. From the growth
of $\delta H$ evident in the plot, one can 
see that the instability sets in between
$\dtau = 0.0105$ and $\dtau =0.0110$. We can also see that the rapid
growth of $\delta H$ is accompanied by a growth in the fermionic forces
in the system (in both norms) and that the $\infty$--norm of the force
appears to grow more rapidly than the degree of freedom averaged 2--norm.
This latter behaviour suggests that the onset of instability is driven
by a few unstable fermion modes, again in line with the 
above hypothesis.

In a further investigation of the MD forces, we carried out MD
trajectories using the same initial gauge configuration as before,
this time varying $\kappa$ for two separate values of the step
size. The values of the step size were $\dtau = 0.010$ and $\dtau =
0.012$ corresponding to stable and unstable MD at $\kappa = 0.1355$
respectively, as discussed above.

\begin{figure}[ht]
\begin{center}
\leavevmode
\hbox{%
\epsfxsize=4.5in
\epsffile{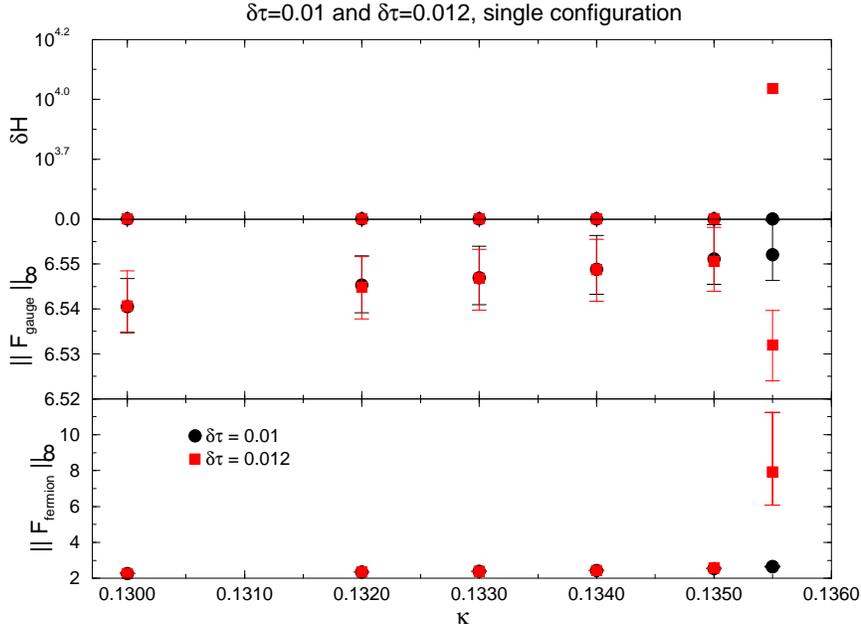}}
\end{center}
\caption{The $\infty$--norms of the gauge and fermionic forces, and 
$\delta H$ against $\kappa$.}
\label{f:Forces_vs_kappa}
\end{figure}

We show the $\infty$--norms of the gauge and fermion forces in figure
\ref{f:Forces_vs_kappa}. This shows that the
simulation which was unstable at $\kappa = 0.1355$ has become stable as
$\kappa$ is reduced.
Once again this seems in line with the hypothesis that the onset of
the instability is a function of the combination of the fermionic
forces (controlled by $\kappa$) and the stepsize
$\dtau$. Recall that the relevant parameter for
the SHO was $\omega\dtau$. 

Overall, our studies of the MD forces lend support to the hypothesis
that the instability is driven by the $F\dtau$ term in the momentum 
update step of the leapfrog algorithm. Since the fermionic force
diverges in some inverse relation with the fermion mass, we expect 
the maximum safe stepsize $\dtau$ to decrease as the 
fermion mass is decreased ($\kappa$ is increased). Also, having observed
a faster rise in the $\infty$--norm of the fermionic force than in 
the degree of freedom averaged 2--norm, we infer that the instability
is driven by a comparatively small number of unstable fermionic modes.

\section{Tuning the stepsize and the solver
residue}\label{s:TuningStudy2}
The above conjecture, if correct, can serve to explain the tuning
results described in section~\ref{s:SolverTuning}.  By increasing the
solver residue $r$, we are modifying the fermionic force which could
then drive the MD integrator unstable.  In order to investigate these
possibilities, we have carried out a second tuning exercise this, time
varying both the step size $\dtau$ and the solver target residue $r$. 

We used the 10 configurations used when tuning $r$ alone in section
\ref{s:SolverTuning}. Since at this point we were not computing
Liapunov exponents, our tests consisted of single MD trajectories in
one direction only. For each value of $\dtau$, we chose the number of
steps along the trajectory so as to maintain a constant trajectory
length of $\tau=175/180$. We also carried out a test with a target
residue of $r = 10^{-9}$ using double precision (64bit) floating point
numbers, whereas all other tests used single precision.  For each
combination of algorithmic parameters, we measured the energy change
$\delta H$, the corresponding acceptance probability $P_{\rm acc}$ and
the cost function of (\ref{e:cost_Func}).

\begin{figure}[ht]
\begin{center}
\leavevmode
\hbox{
\epsfxsize=4.5in
\epsffile{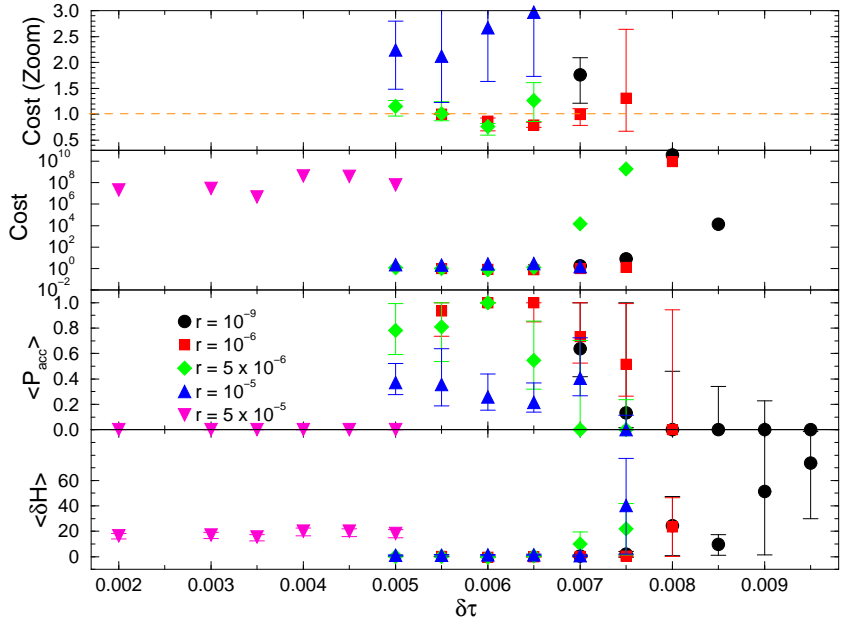}}
\end{center}
\caption{The average energy change $\Expt{\delta H}$, acceptance probability
$\Expt{P_{\rm acc}}$ and cost function for the second tuning study, plotted
against step--size $\dtau$ for several values of the solver target residue 
$r$. The cost function is normalised by its value when $\dtau = 0.0055$,
and $r=10^{-6}$.}
\label{f:all_vs_dtau}
\end{figure}

The results of this tuning exercise are shown in figure
\ref{f:all_vs_dtau}.  First we see in the bottom panel
($r=10^{-9}$ symbols) that using double precision does not alleviate
the problem of instability. The calculation in double
precision appears to become unstable at a similar value
of the step size as does that in single precision. 
Second, we see from the data for $r=5 \times 10^{-5}$ that,
if the solver target
residue is too large, one cannot achieve values of $\delta H$ of
$O(1)$, even if $\dtau$ is made very small.

For our simulations, we are able to achieve non--zero acceptance rates
when $\dtau < 0.0075$ and when $r \le 10^{-5}$. For parameter values
smaller than these, we can attempt to tune our simulation for maximum
performance.  The top two panels of figure \ref{f:all_vs_dtau} show the
variation of the cost function. 
In this case, the cost function is
normalised by its value when $r=10^{-6}$ and $\dtau = 0.0055$.  
These were the parameters used in the production of the dataset from which
the configurations were taken. We see that either by tuning the
solver residue $r$ or the MD step size $\dtau$, the maximum gain we
could make in the cost function is about 25\%.

\section{Conclusions and Discussion} \label{s:Conclusions}
\subsection{Stability} \label{s:CStability}
We have shown that, for the physical
parameters used in our production simulations, 
the molecular dynamics integrator 
used becomes unstable at $\delta
\tau \approx 0.01$ for all studied values of $r$, and also for any
realistic value of $\delta \tau$ when $r$ was increased above $r
\approx O(10^{-5})$. We identify this instability with the one studied 
in free field theory for the frequency--step-size combination $\omega_{\rm
max} \dtau = 2$. We have studied numerically the fermion force and
found that its behaviour is not inconsistent with 
the hypothesis of \cite{AdkHorvath}
(motivated by free field theory) 
that the force should grow large as $\kappa
\rightarrow \kappa_c$.  We suppose that a critical value exists for
$F\dtau$ when the leapfrog integrator becomes unstable.

Reducing the value of the MD residual results in
an increasingly inaccurate
force calculation.
If as a result $F$ is too large, one may need an extremely small step-size to 
keep the integrator stable. We found that, for $r = 5\times 10^{-5}$
at our parameters, one would need a step-size much smaller than $\dtau = 0.001$.
(c.f. figure \ref{f:all_vs_dtau}).

On the safe side of these limits, one may attempt to tune the
algorithm. However, our studies show that on this volume and with these physical
parameters,
tuning $\delta \tau$ and/or $r$  
is unlikely to produce significant performance gains. 
We note that it
appears entirely safe to carry out computations in single precision in
the safer region of parameter space. However, as $\kappa \rightarrow
\kappa_c$, it may be that the upper limit on $r$ decreases beyond the
limit of single precision. Alternatively, as the condition number of
the fermion matrix increases with increasing $\kappa$, the number of
iterations in the solver for fixed $r$ will increase.  This may
cause rounding errors to accumulate so that the target residual $r$ 
may not be reached. 
However, in this latter case, it is only the solve itself that needs
to be done in double precision, or restarted in single precision.

\subsection{Higher order integration Schemes} We have demonstrated
that, at least for the case of a simple harmonic oscillator, the 5th
and 7th order schemes of \cite{Campostrini,Creutz} are not immune to
instabilities. We expect that this situation will persist for even
higher order schemes of this sort. The source of the problem is that,
at the bottom level, these schemes are constructed out of simple
leapfrog updates. For any given step--size $\dtau$ in an integration
scheme of order $n+3$, there will always be a sub update of order
$n+1$ which will have a stepsize $\dtau_2 > \dtau$.  This
sub--update, or one of its constituent sub--updates, may
eventually drive the whole integration scheme unstable, although the
other sub-updates may act as a stabilizing factor at first.  We note
that, in our harmonic oscillator examples, the smallest positive
critical value of $\omega \dtau$ was always smaller for the higher
order integrators than for the leapfrog, indicating that the
instability problem is actually {\em worse} for the higher order
methods.

As the source of the instability appears to come from the
fermionic part of the force, we anticipate that a scheme of the
type advocated in \cite{SextonWeingarten} would not assist avoiding
the instability either, as it attempts to improve the truncation error
by performing more gauge updates. While this may drive down the truncation
error, it does nothing about the problem in the fermionic update.

\subsection{Reversibility} \label{s:CReversibility}
Reversibility itself seems not to be strongly affected by changing
$r$. The Liapunov exponents of the system seem to show a slow rise before the
instability sets in. In the region of transition from stability to
instability, the Liapunov exponents are difficult to determine. 
One might speculate that this behaviour reflects a transition from the
Liapunov exponent characterising the underlying continuous equations
of motion to that characterising the unstable
numerical integrator.

\subsection{Summary}
We have investigated the stability and reversibility of the HMC
algorithm with two flavours of light dynamical fermions on large lattices
as a function of the MD step size $\delta \tau$ and
the MD target solver residue $r$. We have found upper limits on both
of these for a fixed set of physical parameters. Beyond these limits,
the leapfrog integrator becomes unstable and one cannot
carry out a simulation programme, irrespective of the precision of the
floating point numbers which one uses. On the safe side of the limits,
one can carry out simulations safely in both single and double
precision. Parameter tuning seems to give no major performance
gains. Reversibility does not seem to be dangerously affected.

\section{Acknowledgements}
We gratefully acknowledge financial support from PPARC under grant number GR/L22744. James Sexton would like to thank Hitachi Dublin Laboratory for support. We also wish to thank Z. Sroczynski for helpful discussions and for his assistance
in the preparation of this paper.

\newpage